\documentstyle[12pt]{article}
\newcommand{\pabar}{\not{\!\partial}}
\newcommand{\Dbar}{\not{\!{\!D}}}

\title{THE STANDARD MODEL ANOMALIES IN CURVED
SPACE-TIME WITH TORSION}
\author{{\large \bf Antonio Dobado and Antonio L. Maroto} \\
Departamento de F\'{\i}sica Te\'orica \\
Universidad Complutense de Madrid\\
 28040 Madrid, Spain }
\begin{document}
\maketitle
\begin{abstract}
Using the Fujikawa and the heat-kernel methods we make a complete and
detailed computation of the global, gauge and gravitational anomalies
present in the Standard Model defined on a curved space time with
torsion. We find new contributions coming from curvature and torsion
terms to the leptonic number anomaly (so that $B-L$ is not conserved any
more), to the $U(1)_Y$ gauge and to the mixed $U(1)_Y$-gravitational
anomalies, but the gauge anomaly cancellation conditions on the
hypercharges remain the same. We also find that the condition, usually
related to  the cancellation of the mixed $U(1)_Y$-gravitational anomaly,
can be reobtained in the context of the Standard Model
 in flat space-time by requiring the cancellation of the global Lorentz anomaly
without any reference to gravitation.

 \end{abstract}
FT/UCM/8/95

\newpage
\baselineskip 0.83 true cm
\textheight 20 true cm
\newpage

\section{Introduction}

In the last two decades an enormous amount of work has been devoted to the
formulation of different extensions of the Standard Model (SM). These
extensions
include Kaluza-Klein models,  Grand
Unified Theories, Supersymmetry, Supergravity, Superstrings and so on
(different
reviews on these topics can be found in \cite{rev}). The main
goal of these theories is to provide an unified framework for all the known
interactions including
gravitation in many cases. Nevertheless, in spite of the great number of
achievements obtained in
those fields, no general consensus exists on which should be  the most
appropriate approach to the
description of the known interactions. Even in the case of the heterotic
string, which seems to be
the most promising theoretical framework in the opinion of many theoricists,
the low-energy
predictions rely very much in the particular choice of the details of the
compactification of the
extra dimensions which are needed for a proper formulation of (critical) string
theories.

At this point it seems to be interesting to recapitulate and try to go back to
the point where
fundamental physics, understood as a positive science, lies today. From this
point of view,
the amount of knowledge on fundamental interactions confirmed experimentally
can be summarized roughly in the SM, considered as a Quantum Field Theory
(QFT), and Classical Gravitation (CG) by which we mean General Relativity or
other
geometrical theories where the gravitational field is described as a space-time
curvature thus
including the Equivalence Principle (EP). By this we mean that any phenomenon
ever observed can in principle be accommodated in the SM formulated
in a curved space-time background. Of course there are many reasons to think
that
this is not the final theory (provided such a thing exists at all)  but at
least
it is the minimal one compatible with all the experimental data.

For the above reason we consider an important issue the proper formulation of
the SM
in curved space-time. The problem of defining a QFT in curved space-time has
been
considered in detail in the literature some time ago (see \cite{birrel} for a
review). Concerning the particular case of the SM, the most important property
is that it is a chiral gauge theory based on the gauge group $SU(3)_C \times
SU(2)_L \times U(1)_Y$. As the matter (quarks and fermions) is described by
fermionic  fields, one is forced to introduce vierbeins and connections on the
space-time manifold. As it is well known, once a metric is given,
there is only one connection which is metric compatible and torsion free,
namely
the Levi-Civita connection which is defined by the Christoffel symbols. In fact
this was the connection considered by Einstein in his original formulation of
General Relativity. However,  one can also consider the vierbein and the
connection as independent structures. In this case, if one  starts from the
standard Einstein-Hilbert action one gets again the Christoffel symbols for the
connection as a solution of an equation of motion together with the Einstein
field equations for the metric (Palatini formalism). Thus  in this case we also
obtain the Levi-Civita connection on-shell. However quantum effects or
modifications of the action obtained for example by adding higher derivatives
terms to the Einstein-Hilbert action could produce torsion. In addition,
fermions, like the ones appearing in the SM, give a non-zero contribution to
the
torsion (see for example \cite{Hehl}). Finally, most of the extensions of
General Relativity introduce the vierbein and the connection as independent
entities and this will be our approach in the following.
Nevertheless, we will keep the covariant constancy
condition for the metric, in order to have a geometrical meaning for the
connection. This condition amounts to consider the connection as a $SO(4)$ or
$SO(3,1)$ Lie algebra valued one form (for Euclidean or Lorentzian signature
respectively).

Thus in the following we will address the problem of defining properly the SM
as a QFT in
presence of a classical space-time with torsion. As it is well known theories
with chiral
fermions like the SM are potentially plagued of gauge and gravitational
anomalies which can
ruin the consistency of the quantum theory even if it is well defined at the
classical level. Fortunately, the current assignment of hypercharges for the
different fermions appearing in the SM is done in such a subtle way that all
those anomalies exactly cancel. In addition we have also anomalies affecting
some other global classical symmetries that can give rise to interesting
physical effects like the non-conservation of the baryonic or leptonic numbers.

In this work we will compute all those anomalies in the SM defined on a curved
space-time with
torsion. As we will see, the torsion will give new contributions to most of
those gauge, gravitational and global anomalies which are far from trivial. The
plan of the paper goes as follows: In section two we develop the formalism for
the definition of the SM in a curved background space-time with torsion. In
section three we discuss the technicalities for the computation of the
different
anomalies including the appropriate versions of the Fujikawa and heat-kernel
methods. In section four we consider the baryon and lepton anomalies in
presence
of torsion. In section five we compute the anomalies of the gauge  $SU(3)_C
\times SU(2)_L \times U(1)_Y$ group currents including the new torsion
contributions. In section
six we study the gravitational anomalies as anomalies in the Lorentz group
understood as a gauge
(local) group.
In section seven we discuss some of the consequences of our previous anomaly
computation and in
particular those concerning the quantization of the electric charge in the
framework of the SM. Finally, in section eight we briefly list the main
conclusions of our work.

\section{The Standard Model in curved space-time with torsion}

The formulation of the SM interacting with classical gravity is based on the
Einstein Equivalence
Principle (EP) (see for example
\cite{Ramond}). The EP makes it possible to obtain the
 curved space-time Physics (i.e. General Relativity) from that of the
flat space-time (i.e. Special Relativity), since it determines the nature of
the interaction
 with gravity of any other field. This principle states that at
each point $p$ of
space-time it is always possible to find a privileged coordinate system in
which physics
looks locally like in flat space-time. The procedure one should follow in order
to introduce the
gravitational interaction in any field theory built in flat space-time goes as
follows: take the
Lorentz invariant action of
 the
theory and identify the coordinates appearing in it with that of the locally
inertial system. Then
 perform
a coordinate change to an arbitrary coordinate system and the gravitational
interaction will appear
automatically. As in  this work we are mainly interested in the effect of
gravitation on anomalies,
we will start by applying this recipe in detail to work out the gravitational
interaction of Dirac
spinors. At
 the end we will obtain also the lagrangians
for scalar and gauge fields interacting with gravity.

Let us first introduce some notation. We will use latin indices $m,n...$ for
objects referred to the locally inertial coordinate system and Greek indices
$\mu,\nu...$ for any other. If $\{ \xi^m \}$ are the coordinates in the
privileged system and $\{ x^{\mu}\}$ the coordinates in any other, then:

\begin{eqnarray}
g^{\mu \nu}(x)=e_m^{\mu}(x)e_n^{\nu}(x)\eta^{mn}
\end{eqnarray}
where $\eta^{mn}=(-,-,-,-)$ is the Euclidean flat metric once the Wick
rotation has been done (as usual in functional calculations we will work in
Euclidean space): $x^0\rightarrow -i\hat x^4$, $x^i \rightarrow \hat x^i$,
$\partial_0 \rightarrow i\hat \partial_4$, $\partial_i \rightarrow
 \hat \partial_i$ and the
 Euclidean gamma matrices: $\gamma^0\rightarrow \hat \gamma^4$,
$\gamma^i \rightarrow i\hat\gamma^i$. We define: $\hat \gamma_5=
-\hat \gamma_1\hat \gamma_2\hat \gamma_3\hat \gamma_4$.
 The vierbein $e_m^{\mu}(x)=\partial x^{\mu}/\partial {\xi^m}$
gives at each point the change of coordinates to the
privileged system. Analogously it is possible to define  an inverse vierbein by
$e_m^{\mu}e^n_{\mu}=\delta _m^n$ and $e_m^{\mu}e^m_{\nu}= \delta ^{\mu}_{\nu}$.
Finally let us introduce the volume form written in terms of vierbeins:
\begin{eqnarray} d^4\xi=\sqrt{g}d^4x=(\det e^m_{\mu})d^4x
\end{eqnarray}
with $g=|\det g_{\mu\nu}|$.

In flat space-time Dirac spinors change in the following way under
Lorentz transformations:
\begin{eqnarray}
\psi(p)  & \rightarrow &
U\psi(p)=e^{\frac{i}{2}\epsilon^{mn}\Sigma_{mn}}\psi(p) \nonumber \\
\psi^{\dagger}(p) & \rightarrow &
\psi^{\dagger}(p)U^{\dagger}=\psi^{\dagger}(p)e^{-\frac{i}{2}\epsilon^{mn}\Sigma_{mn}}
\end{eqnarray}
where $\Sigma_{mn}=\frac{i}{4}[\gamma_m,\gamma_n]$ are the hermitian
generators of the $SO(4)$ group in the spinor representation.

The Dirac lagrangian in flat space-time

\begin{equation}
{\cal L}=\frac{1}{2}(\psi^{\dagger}\pabar \psi-\partial_{m}\psi^{\dagger}
\gamma^{m}
\psi)
\label{dirac}
\end{equation}
 is invariant under those global transformations. Notice that we have
written the hermitian form of the lagrangian in Euclidean space and with
fermions considered as anticommuting variables. In flat space-time it is
always possible to integrate by parts and write the lagrangian in the
more usual way:

\begin{eqnarray}
{\cal L}=\psi^{\dagger} \pabar \psi
\end{eqnarray}

Now, the EP  requires this invariance of the Dirac
lagrangian under Lorentz transformations to be not only global but also local
when
gravitation is included. This fact forces us to introduce a covariant
derivative
for this gauge  transformation in order to make eq.\ref{dirac} invariant.
Therefore, let us write the gauged hermitian Dirac lagrangian in the following
way:

\begin{eqnarray}
{\cal L}=\frac{1}{2}(\psi^{\dagger}\gamma^m D_m\psi-D_m \psi^{\dagger}\gamma^m
\psi)
\end{eqnarray}

The EP  has allowed us to write the Dirac
lagrangian in the privileged system. Now we can write it in any other
coordinate system
 by using the vierbein:

\begin{eqnarray}
{\cal L}=\frac {1}{2} (\psi^{\dagger} \gamma^{\mu}D_{\mu}\psi-D_{\mu}
\psi^{\dagger}\gamma^{\mu}\psi)
\end{eqnarray}
where we have introduced the Dirac matrices in curved space-time
$\gamma^{\mu}=e^{\mu}_m \gamma^m$. These matrices satisfy a similar
algebra in curved space-time: $\{\gamma^{\mu},\gamma^{\nu}\}=-2g^{\mu\nu}$. The
covariant derivative
 is defined as usual by:

\begin{eqnarray}
D_{\mu}=(\partial_{\mu}+\Omega_{\mu})
\end{eqnarray}
where $\Omega_{\mu}$ is known as the spin connection. The transformations
rules of $\Omega_{\mu}$ under local Lorentz transformations are those of
a gauge connection:

\begin{eqnarray}
\Omega_{\mu}\rightarrow \Omega_{\mu}'=U(x)\Omega_{\mu}U^{-1}(x)-
(\partial_{\mu}U)U^{-1}(x)
\end{eqnarray}
or infinitesimally:

\begin{eqnarray}
\Omega_{\mu}\rightarrow \Omega_{\mu}+\frac{i}{2}\epsilon^{ab}(x)
[\Sigma_{ab},\Omega_{\mu}]-\frac{i}{2}(\partial_{\mu}\epsilon^{ab}(x))
\Sigma_{ab}
\end{eqnarray}

Now, recalling that the components of the connection 1-form in Riemannian
geometry have precisely the latter transformation rule \cite{Nakahara}, we can
identify:

\begin{eqnarray}
\Omega_{\mu}=-\frac{i}{2}\hat\Gamma^{a\;b}_{\;\mu}\Sigma_{ab}
\end{eqnarray}
and define the covariant derivative acting on spinors as:

\begin{eqnarray}
D_{\mu}\psi=(\partial_{\mu}-\frac{i}{2}\hat\Gamma^{a\;b}_{\;\mu}\Sigma_{ab})\psi
\end{eqnarray}

Depending on the object this derivative acts on, the generators will appear
in the corresponding representation (vector, tensor, etc) of the Lorentz
group. It is easy to see that this gauge covariant derivative is nothing
but the ordinary geometric covariant derivative but referred to the
privileged coordinate system. However, this gauge formulation of the Lorentz
group enables to introduce spinors in curved space-time which otherwise
would be impossible, since $GL(4)$ does not posses spinor representations.

Notice that $\{\hat\Gamma^{a\;b}_{\mu}\}$ does not have to be
a torsion free Levi-Civita connection, which we will denote
$\{\Gamma^{a\;b}_{\;\mu}\}$.
In general, if we take a connection compatible with the metric, i.e.
$\hat\nabla_{\nu}g_{\alpha \beta}=0$, then it can be written as:

\begin{eqnarray}
\hat\Gamma^{a\;b}_{\;\mu}=\Gamma^{a\;b}_{\;\mu}+
e^a_{\nu}e^{\lambda b}K^{\nu}_{\;\mu\lambda}
\label{tors}
\end{eqnarray}
where $K^{\nu}_{\;\mu\lambda}$ is known as the contorsion tensor which
 in terms of the torsion tensor reads \cite{Nakahara}:

\begin{eqnarray}
 K^{\nu\mu\lambda}=\frac{1}{2}(T^{\nu\mu\lambda}+T^{\mu\nu\lambda}+
T^{\lambda\nu\mu})
\end{eqnarray}

Now, using the decomposition  in eq.\ref{tors} we can write the Dirac
lagrangian with an arbitrary metric connection in terms of the usual
Levi-Civita plus
 an additional
 term
depending on the torsion \cite{Buch}:

\begin{eqnarray}
{\cal L}=\frac{1}{2}(\psi^{\dagger}\gamma^{\mu}(\partial_{\mu}-\frac{i}{2}
\hat\Gamma^{a\;b}_{\mu}\Sigma_{ab})\psi-(\partial_{\mu}\psi^{\dagger}+\frac{i}{2}
\hat\Gamma^{a\;b}_{\mu}\psi^{\dagger}\Sigma_{ab})\gamma^{\mu}\psi)\nonumber \\
=\psi^{\dagger}\gamma^{\mu}(\partial_{\mu}-\frac{i}{2}
\hat\Gamma^{a\;b}_{\mu}\Sigma_{ab}+\frac{1}{2}T_{\mu})\psi=
\psi^{\dagger}\gamma^{\mu}(\partial_{\mu}-\frac{i}{2}
\Gamma^{a\;b}_{\mu}\Sigma_{ab}-\frac{1}{8}S_{\mu}\gamma_5)\psi
\end{eqnarray}
where:

\begin{eqnarray}
S_{\alpha} & = & \epsilon_{\mu\nu\lambda\alpha}T^{\mu\nu\lambda}\nonumber \\
T_{\mu} & = & T^{\lambda}_{\;\lambda\mu}=K^{\lambda}_{\;\lambda\mu}
\end{eqnarray}

Note that with this definition $S_\mu$ is axial the part of the torsion tensor.
In conclusion, the lagrangian for Dirac fermions in a curved space-time with
torsion is that of a fermion in a curved space-time without torsion plus an
axial interaction with $S_\mu$. Nevertheless, there is a difference between the
axial coupling of torsion with the usual axial couplings of gauge fields. While
the latter breaks the hermiticity of the Dirac operator, the former does not.
This similarity  will simplify the computation of the anomalies when using the
well-known heat kernel expansion in curved space-time.

Notice however that this is the minimal lagrangian. If we had used instead a
more general form, some
 other non-minimal couplings with torsion could also have appeared, as for
instance:
$i\psi^{\dagger}\gamma^{\mu} T_{\mu}\psi$ \cite{Buch}. Such non-minimal
coupling
can be discarded by anomaly cancellation  arguments. In fact, this term behaves
as an hypercharge field which interacts with the left and right
 components
with the same coupling constant (except for the neutrino). We will see that
such coupling
yields an anomaly in the $SU(2)_L$, $U(1)_Y$ as well as in the local Lorentz
symmetry.
 As there is no hypercharge assignment that could cancel simultaneously all the
anomalous
contributions,  we will not consider it here.

Now that we know the general expression for the Dirac lagrangian
in a curved space-time with torsion, let us apply it to the SM matter sector,
 which can be written in the following way in the case of massless fermions and
without considering the Yukawa couplings to the Higgs field:

\begin{eqnarray}
{\cal L}_m=Q^{\dagger}{\Dbar_Q}Q+L^{\dagger}{\Dbar_L}L
\label{lm}
\end{eqnarray}
where the Dirac operators for quarks and leptons are defined as:

\begin{eqnarray}
i\Dbar_Q & = & i\gamma^{\mu}(\partial_{\mu}+{\bf \Omega}_{\mu}+{\bf
G}_{\mu}+{\bf W}_{\mu}P_L+
{\bf B}_{\mu}+{\bf S}_{\mu}\gamma_5) \nonumber \\
i\Dbar_L & = & i\gamma^{\mu}(\partial_{\mu}+{\bf \Omega}_{\mu}+{\bf
W}_{\mu}P_L+
{\bf B}_{\mu}+{\bf S}_{\mu}\gamma_5)
 \end{eqnarray}

Here we have organized the matter fields in doublets, so that for the first
family we
 have:

\begin{eqnarray}
Q=\left[
\begin{array}{c}
u\\
d
\end{array}
\right]
\;L=\left[
\begin{array}{c}
\nu\\
e
\end{array}
\right]
\end{eqnarray}

Their left components $Q_L$ and $L_L$ are $SU(2)_L$ doublets, while each
component of the
right part  $Q_R$ and $L_R$ are $SU(2)_L$ singlets. In turn the $u$ and $d$
quarks are
$SU(3)_c$ triplets.  The gauge fields appearing in the operators are:

\begin{itemize}

\item Gluons; which are those corresponding to the $SU(3)_c$ group, that we
will denote by
\begin{eqnarray}
{\bf G}_{\mu}=-ig_SG_{\mu}^a\Lambda^a
\end{eqnarray}
 Here the $\Lambda^a$ are the eight group generators in a product
representation: $\Lambda^a=\lambda^a \otimes 1_2$, where $\lambda^a$ are the
properly normalized
 Gell-Mann matrices and $1_2$ is the $2\times 2$ identity matrix in flavor
space.

\item $W$-bosons, which are those corresponding to the $SU(2)_L$ symmetry that
we will write as
 \begin{eqnarray}
{\bf W}_{\mu}=-igW_{\mu}^aT^a
\end{eqnarray}
where $T^a$ are the three group generators in the product representation:
$T^a=1_3 \otimes \sigma^a/2$ for quarks and $T^a=\sigma^a/2$ for leptons,
 with $\sigma^a$ the Pauli matrices.

\item Finally there is also the hypercharge boson
\begin{eqnarray}
{\bf B}_{\mu}=ig'B_{\mu}\left(P_L \left(
\begin{array}{cc}
y_L^u & \; \\
\; & y_L^d
\end{array}
\right)+P_R\left(
\begin{array}{cc}
y_R^u & \; \\
\; & y_R^d
\end{array}
\right) \right)
\label{hy}
\end{eqnarray}
which  has been written for the case of quarks. For leptons
 the expression is obtained using their corresponding hypercharges.

\end{itemize}

Before commenting on the curvature and torsion terms we should stress that
these operators are
not hermitian. This is
due to the chiral couplings of $SU(2)_L$ and hypercharge fields. Thus the
adjoint operators are:

\begin{eqnarray}
(i\Dbar_Q)^{\dagger} & = & i\gamma^{\mu}(\partial_{\mu}+{\bf \Omega}_{\mu}+{\bf
G}_{\mu}+{\bf
W}_{\mu}P_R+ \bar {\bf B}_{\mu}+{\bf S}_{\mu}\gamma_5) \nonumber \\
(i\Dbar_L)^{\dagger} & = & i\gamma^{\mu}(\partial_{\mu}+\bar{\bf
\Omega}_{\mu}+{\bf W}_{\mu}P_R+
\bar{\bf B}_{\mu}+\bar{\bf S}_{\mu}\gamma_5)
 \end{eqnarray}
where:
\begin{eqnarray}
\bar{\bf B}_{\mu}=ig'B_{\mu}\left(P_R \left(
\begin{array}{cc}
y_L^u & \; \\
\; & y_L^d
\end{array}
\right)+P_L\left(
\begin{array}{cc}
y_R^u & \; \\
\; & y_R^d
\end{array}
\right) \right)
\end{eqnarray}
and analogously for leptons. Notice that, since there is no right neutrino, the
spin connection can be written as follows for leptonic operators:

\begin{eqnarray}
{\bf \Omega}_{\mu}=-\frac{i}{2}\Gamma^{a\;b}_{\mu}\left(
\begin{array}{cc}
P_L\Sigma_{ab} & \; \\
\; & \Sigma_{ab}
\end{array}
\right)
,\; \bar{\bf \Omega}_{\mu}=-\frac{i}{2}\Gamma^{a\;b}_{\mu}\left(
\begin{array}{cc}
P_R\Sigma_{ab} & \; \\
\; & \Sigma_{ab}
\end{array}
\right)
\end{eqnarray}
for the same reason,  the torsion terms are:

\begin{eqnarray}
{\bf S}_{\mu}\gamma_5=-\frac{1}{8}S_{\mu}\left(
\begin{array}{cc}
P_L\gamma_5 & \; \\
\; &\gamma_5
\end{array}
\right)
,\; \bar{\bf S}_{\mu}\gamma_5=-\frac{1}{8}S_{\mu}\left(
\begin{array}{cc}
P_R\gamma_5 & \; \\
\; &\gamma_5
\end{array}
\right)
\end{eqnarray}

Once we have obtained the lagrangian for Dirac spinors in curved space-time
with torsion, we
 will write the corresponding one for scalar fields. The standard lagrangian
for
a scalar field
 in flat space-time can be
written as follows:

\begin{eqnarray}
{\cal
L}=\frac{1}{2}\partial_{\mu}\phi\partial^{\mu}\phi-\frac{1}{2}m\phi^2-V(\phi)
\end{eqnarray}
(we
will consider only the case of a real scalar field since the complex case is
completely analogous).
According to the given prescription to build field theories interacting with
gravity from the EP, we
 have to identify the coordinates in the
lagrangian density with those of the privileged system, and then make
local the Lorentz invariance. As the fields are scalars, they do not change
under Lorentz
 transformations and their covariant derivative is just an ordinary derivative.
Finally we have to
use the vierbein to transform to an arbitrary coordinate system.  Then the
final expression for the
action integral reads

\begin{eqnarray}
S=\int d^4x \sqrt{g}\left ( \frac{1}{2}g^{\mu
\nu}\partial_{\mu}\phi\partial_{\nu}\phi-\frac{1} {2}m\phi^2-V(\phi) \right )
\label{min}
\end{eqnarray}

which is  the minimal coupling to gravity. However  the most general lagrangian
for a scalar field
 interacting with gravity,
up to a given order in derivatives,
could also include terms like $R\phi^2$, $T_{\mu}T^{\mu}\phi^2$,
$q_{\mu\nu\alpha}q^{\mu\nu\alpha}
\phi^2$,
 $\nabla_{\mu}T^{\mu}\phi^2$, etc, where we
have defined:

\begin{eqnarray}
T_{\alpha\beta\mu}=q_{\alpha\beta\mu}+\frac{1}{3}(T_{\beta}g_{\alpha\mu}-
T_{\mu}g_{\alpha\beta})-\frac{1}{6}S^{\nu}\epsilon_{\alpha\beta\mu\nu}
\end{eqnarray}
and $q_{\alpha\beta\mu}$ satisfies
$\epsilon^{\alpha\beta\mu\nu}q_{\alpha\beta\mu}=0$.

Nevertheless, even if we had started with the minimal lagrangian in
eq.\ref{min}, some of these
 terms would appear in the renormalization procedure as one-loop counterterms.
In the case of scalar
 fields interacting with some gauge fields, we should turn the  derivatives in
eq.\ref{min} into gauge
 covariant derivatives, as it happens for instance for the Higgs fields in the
SM.

Once we have worked out the lagrangian for scalars we turn to the gauge
fields. The Yang-Mills lagrangian is given by:

\begin{eqnarray}
{\cal L}=\frac{1}{4}F^a_{mn}F_a^{mn}
\end{eqnarray}

We consider the strength tensor $F_{mn}$ as defined in a locally inertial
coordinate system.  $F_{mn}$ is a Lorentz tensor and $F^a_{mn}F_a^{mn}$ is
invariant under global and local Lorentz transformations. Therefore we only
have to transform it to an arbitrary coordinate system by means of the
vierbein:

\begin{eqnarray}
F^a_{\mu\nu}=e^m_{\; \mu}e^n_{\; \nu}F^a_{mn}
\end{eqnarray}
and the action integral reads:

\begin{eqnarray}
S=\int
d^4x\sqrt{g}\frac{1}{4}g^{\mu\alpha}g^{\nu\beta}F_{\mu\nu}F_{\alpha\beta}
\end{eqnarray} Therefore, the coupling to gravity occurs only through the
vierbein and the spin connection does not appear. Then there is no coupling
between gauge fields and torsion. As
 a consequence, the only fields which may potentially couple  to torsion are
the fermion fields.
It is also important to notice that this is the only way to preserve gauge
invariance when passing
 from flat to curved space-time. In fact the other natural possibility, which
consists in defining

\begin{eqnarray}
F_{\mu\nu}=\hat\nabla_{\mu}A_{\nu}-\hat\nabla_{\nu}A_{\mu}+[A_{\mu},A_{\nu}]
\end{eqnarray}
drives to a Yang-Mills action that is not gauge invariant when torsion is
present, since:

\begin{eqnarray}
F_{\mu\nu} & = &
\hat\nabla_{\mu}A_{\nu}-\hat\nabla_{\nu}A_{\mu}+[A_{\mu},A_{\nu}]
\nonumber \\    & = &
(\partial_{\mu}A_{\nu}-\Gamma^{\kappa}_{\; \mu \nu}A_{\kappa})
-(\partial_{\nu}A_{\mu}-
\Gamma^{\kappa}_{\; \nu \mu}A_{\kappa}) +[A_{\mu},A_{\nu}]  \nonumber \\
 & = &
(\partial_{\mu}A_{\nu}-\partial_{\nu}A_{\mu}+[A_{\mu},A_{\nu}])-T^{\kappa}_{\;
\mu\nu}A_{\kappa} \end{eqnarray}
Here the  first term  is gauge covariant, whereas the last term explicitly
breaks the gauge invariance in the Yang-Mills action for
 non-vanishing torsion.

\section{The heat-kernel for the Standard Model operators}

In this section we will study the possible violation, due to quantum effects,
of the SM classical symmetries, which are of two kinds:

\begin{itemize}

\item  Those which are exact, that  can either be gauge symmetries as
$SU(3)_c$,
 $SU(2)_L$, $U(1)_Y$
  and
the Lorentz group, or global as those corresponding to the lepton and
baryon number conservation.

\item There are also those which are approximate which we will not
consider here. Some examples are the $U(1)_A$, the  $SU(3)_L \times SU(3)_R$
chiral symmetries in
low-energy  QCD and the $SU(2)_L \times SU(2)_R$ global symmetry of the
symmetry breaking sector of the
SM which are only  exact in certain limits.
\end{itemize}

The non-conservation of the gauge symmetries
due to anomalies leads to the inconsistency of the model. Therefore, it is
interesting that the
 inclusion of the
gravitational interaction does not affect the anomaly cancellation in gauge
currents. In addition, the
 gravitational contribution to the non conservation of lepton and baryon
number, could have some
relevance concerning to the problem of  the baryon number asymmetry of the
Universe. Let us then
discuss the status of each of these symmetries at the quantum level in a curved
space-time with
torsion.

There are several techniques proposed for the computation of anomalies in the
literature. For our
purposes here the most appropriate is to use
the functional methods  that were first introduced by Fujikawa
\cite{Fujikawa} in a flat space-time and later
 extended to curved space-time by Yajima \cite{Yajima}. According to these
methods we need an
hermitian operator in order to regularize the anomalous path integral jacobian
of the symmetry
transformation. Thus for instance, for the axial anomaly the transformation
yields:

\begin{eqnarray}
[d\psi d\psi^{\dagger}]\rightarrow[d\psi d\psi^{\dagger}]
exp(-2\int d^4x\sqrt {g} i\alpha (x) A(x))
\end{eqnarray}
where the anomaly $A(x)$ appearing in the (regularized) jacobian reads
\begin{eqnarray}
A_{reg}=\lim_{t \rightarrow \infty} \sum_n \phi_n^{\dagger}\gamma_5
e^{-\frac{\lambda^2}{t^2}}\phi_n=
\lim_{t \rightarrow \infty} \sum_n \phi_n^{\dagger}\gamma_5
e^{-\frac{(i\Dbar)^2}{t^2}}\phi_n=
\lim_{t \rightarrow \infty}
tr \frac{t^2}{(4\pi)^2}\gamma_5\sum_{n=0}^{\infty} \frac{a_n(x)}{t^n}
\end{eqnarray}
where $\lambda_n$ are the eigenvalues of the hermitian operator of the theory
and in the last
step we have performed a heat-kernel expansion. In general, the expression
above is divergent, in
 the
$t \rightarrow \infty$ limit, due to the two  first terms in the
heat-kernel expansion. In such case, certain renormalization prescription
 will be needed to obtain a finite value for the anomaly. However, it may
happen, as it occurs in
 theories with only vector couplings to gauge fields, that those potentially
divergent terms vanish
and $A_{reg}$ is finite.

One prescription to eliminate the divergent terms in the anomaly consists in
removing them directly.
 This
drastic procedure can be justified in some circumstances as follows
\cite{Gamboa}. Let us define the
 transformation jacobian as the quotient between the effective action and the
transformed effective
action, both regularized using $\zeta$-function regularization. Thus, for
instance, in the case of
 the axial transformations considered before:

\begin{eqnarray}
det \Dbar=J det(e^{\gamma_5 \alpha(x)}\Dbar e^{\gamma_5 \alpha(x)})
\end{eqnarray}
 where $J$ is the jacobian of the symmetry transformation. It is then possible
to show that, provided
 the operator is hermitian, the result for the anomaly is the same
 as the one obtained using the Fujikawa
method and removing the divergent terms, i.e: $A=\frac{1}{(4\pi)^2}tr \gamma_5
a_2(x)$. Therefore,
 this is the
prescription we will use to render the result finite. However, for  that
purpose, we need an
hermitian
 operator as a
regulator but, as we have seen, the Dirac operator in the SM is not hermitian.
Several methods
 have been
proposed \cite{Fujikawa2} to avoid this problem, we will mention two of them.
 In the first one we split the
lagrangian and the integration measure in their left and right components:

\begin{eqnarray}
[d\psi d\psi^{\dagger}]\rightarrow
[d\psi_R^{\dagger}d\psi_L^{\dagger}d\psi_Rd\psi_L]\nonumber \\
\psi^{\dagger}\Dbar\psi=\psi^{\dagger}_R\Dbar_L\psi_L+\psi^{\dagger}_L\Dbar_R\psi_R
\end{eqnarray}

In curved space-time without torsion (the torsion term is written between
brackets)
 the  Dirac quark operators

\begin{eqnarray}
i\Dbar_L & = & i\gamma^{\mu}(\partial_{\mu}+{\bf \Omega}_{\mu}+{\bf
G}_{\mu}+{\bf W}_{\mu}+
 {\bf B}_{\mu}^L (-{\bf S}_{\mu})) \nonumber \\
i\Dbar_R & = & i\gamma^{\mu}(\partial_{\mu}+{\bf \Omega}_{\mu}+{\bf G}_{\mu}+
 {\bf B}_{\mu}^R(+{\bf S}_{\mu}))
\end{eqnarray}
are hermitian (the same is true for leptons). Thus they allow the
regularization of the corresponding
 piece
of the anomaly. However, the torsion term  breaks the hermiticity of these
operators and therefore
 this
method does not seem to be suitable in presence of torsion. In spite of this
fact, one could rotate
 $S_{\mu} \rightarrow iS_{\mu}$ \cite{Adrianov}. This makes the operators in
the above equations
hermitian and then, at the end of the calculation, one can undo the rotation.
Such procedure has been
proved to be
useful in theories with axial gauge couplings and yields the so called
consistent
 anomaly.
Notice however that, in presence of torsion, certain inconsistencies appear,
since it can be shown that there would not
 be
any  appropriate choice of hypercharges in the SM that could cancel the gauge
anomalies.

An alternative method \cite{Fujikawa2} which does not suffer from this
inconsistencies is to
 regularize  separately those pieces in the anomaly coming from the
transformation of $\psi$ and
${\psi}^{\dagger}$. In this case our first step is to build two hermitian
operators which preserve all
the gauge symmetries in the lagrangian, namely:

\begin{eqnarray}
H_{\psi} & = & (i\Dbar)^{\dagger} (i\Dbar) \nonumber \\
H_{{\psi}^{\dagger}} & = & (i\Dbar)(i\Dbar)^{\dagger}
\label{op}
\end{eqnarray}

 Then the hermiticity ensures that their corresponding eigenfunctions form a
complete
set:
\begin{eqnarray}
H_{\psi}\phi_n & = & \lambda_n^2\phi_n \nonumber \\
H_{\psi^{\dagger}}\xi_n & = & \lambda_n^2\xi_n
\label{aut}
 \end{eqnarray}

 Now we expand $\psi$ and $\psi^{\dagger}$ in terms of eigenfunctions of
$H_{\psi}$ and $H_{{\psi}^{\dagger}}$
 respectively:
\begin{eqnarray}
\psi & = & \sum_{n} a_n \phi_n \nonumber \\
\psi^{\dagger} & = & \sum_{n} \bar b_n \xi_n^{\dagger}
\end{eqnarray}

Under an infinitesimal transformation like (the
 non-abelian
case follows the same steps):
\begin{eqnarray}
\psi  & \rightarrow & \psi+i\alpha (x) \psi \nonumber \\
\psi^{\dagger}  & \rightarrow & \psi^{\dagger} -i\psi^{\dagger} \alpha (x)
\end{eqnarray}
the integration measure changes as:
\begin{eqnarray}
[d\psi d\psi^{\dagger}]\propto [da_n d\bar b _n]\rightarrow [da'_nd \bar b_n']=
[da_n d\bar b _n]
exp(-\int d^4x\sqrt {g} i\alpha (x) A(x))
\label{med}
\end{eqnarray}
where, in the present case,  $A(x)$ is the vector abelian anomaly
\begin{eqnarray}
A(x)=\sum_{n} \phi^{\dagger}_n \phi_n -\sum_{n}\xi^{\dagger}_n \xi_n
\end{eqnarray}

As it has already been mentioned, we regularize each piece of the anomaly with
the corresponding
 operator:

\begin{eqnarray}
A(x)=\lim_{M \rightarrow \infty} \sum_{n} \phi^{\dagger}_n
e^{-(\frac{H_\psi}{M})^2}\phi_n -\sum_{n}\xi^{\dagger}_n
e^{-(\frac{H_{\psi^{\dagger}}}{M})^2} \xi_n
\end{eqnarray}

In order to obtain a finite result, we have to perform the heat-kernel
expansion for the $H_{\psi}$
 and
$H_{\psi^{\dagger}}$ operators and subtract the divergent terms. However,
still a new difficulty
appears. Although the heat-kernel expansion has been worked out for a wide
class of operators
even in curved space-time, the
coefficients become unmanaegable \cite{yan} for operators which do not cast the
general form:

\begin{eqnarray}
H=D_{\mu}D^{\mu}+X
\end{eqnarray}
where $X$ does not contain derivatives. At first glance, this is not the case
of  $H_{\psi}$ and
 $H_{\psi^{\dagger}}$. However,
with some algebra we can write them in the desired form \cite{Cognola}:

\begin{eqnarray}
H_{\psi}=(i\Dbar)^{\dagger}(i\Dbar)=D_{\mu}D^{\mu}+\gamma_5{\bf S}^{\mu}_{\;
;\mu}+2{\bf S}_{\mu}
{\bf S}^{\mu}-
\frac{1}{4}[\gamma^{\mu},\gamma^{\nu}][d_{\mu},d_{\nu}] \nonumber \\
H_{\psi^{\dagger}}=(i{\Dbar})(i{\Dbar})^{\dagger}=\bar D_{\mu} \bar
D^{\mu}+\gamma_5\bar{\bf S}^{\mu}_{\; ;\mu}+
2\bar{\bf S}_{\mu}\bar{\bf S}^{\mu}-
\frac{1}{4}[\gamma^{\mu},\gamma^{\nu}][\bar d_{\mu},\bar d_{\nu}]
\label{canon}
\end{eqnarray}
where for quarks:

\begin{eqnarray}
d_{\mu}=\partial_{\mu}+{\bf {\Omega}}_{\mu}+{\bf G}_{\mu}+{\bf W}_{\mu}P_L+{\bf
B}_{\mu} \nonumber
 \\
\bar d_{\mu}=\partial_{\mu}+{\bf {\Omega}}_{\mu}+{\bf G}_{\mu}+{\bf
W}_{\mu}P_R+\bar{\bf B}_{\mu}
\end{eqnarray}
and
\begin{eqnarray}
D_{\mu}=d_{\mu}-\frac{1}{2}\gamma_5[\gamma^{\mu},\gamma^{\nu}]{\bf S}^{\nu}
\nonumber \\
\bar D_{\mu}=\bar d_{\mu}-\frac{1}{2}\gamma_5[\gamma^{\mu},\gamma^{\nu}]{\bf
S}^{\nu}
\end{eqnarray}

In the case of leptons we have:

\begin{eqnarray}
d_{\mu}=\partial_{\mu}+{\bf {\Omega}}_{\mu}+{\bf W}_{\mu}P_L+{\bf B}_{\mu}
\nonumber \\
\bar d_{\mu}=\partial_{\mu}+\bar{\bf {\Omega}}_{\mu}+{\bf W}_{\mu}P_R+\bar{\bf
B}_{\mu}
\end{eqnarray}
and:

\begin{eqnarray}
D_{\mu}=d_{\mu}-\frac{1}{2}\gamma_5[\gamma^{\mu},\gamma^{\nu}]{\bf S}^{\nu}
\nonumber \\
\bar D_{\mu}=\bar
d_{\mu}-\frac{1}{2}\gamma_5[\gamma^{\mu},\gamma^{\nu}]\bar{\bf S}^{\nu}
\end{eqnarray}
where $\bar{\bf B}_{\mu}$,$\bar{\bf {\Omega}}_{\mu}$ and $\bar{\bf S}^{\nu}$
have been defined
 above. Therefore
 we have already an appropriate form to use the heat-kernel expansion. Now
removing the divergent
 $a_1(x)$
coefficient we obtain for the anomaly in  the abelian vector currents:

\begin{eqnarray}
A(x)=\frac{1}{(4\pi)^2} tr(a_2(H_{\psi},x)-a_2(H_{\psi^{\dagger}},x))
\label{ar}
\end{eqnarray}
where the second coefficient in the heat-kernel expansion in curved space-time
has been worked out in different references
\cite{Yajima}\cite{yan}\cite{Cognola} using different methods
and in our case reads:

\begin{eqnarray}
a_2(H_{\psi},x)=\frac{1}{12}[D_{\mu},D_{\nu}][D^{\mu},D^{\nu}]+\frac{1}{6}[D_{\mu},[D^{\mu},X]]+
\frac{1}{2}X^2-
\frac{1}{6}RX \nonumber \\
-\frac{1}{30}R_{;\mu}^{\;\;\mu}+\frac{1}{72}R^2+\frac{1}{180}(R_{\mu \nu \rho
\sigma}  R^{\mu \nu \rho \sigma} -R_{\mu \nu} R^{\mu \nu})
\end{eqnarray}
and:
\begin{eqnarray}
a_2(H_{\psi^{\dagger}},x)=\frac{1}{12}[\bar D_{\mu},\bar D_{\nu}][\bar
D^{\mu},\bar D^{\nu}]+\frac{1}{6}
[\bar D_{\mu},[\bar D^{\mu},\bar X]]+\frac{1}{2}\bar X^2-
\frac{1}{6}\bar R \bar X \nonumber \\
-\frac{1}{30}\bar R_{;\mu}^{\;\;\mu}+\frac{1}{72}\bar R^2+\frac{1}{180}(\bar
R_{\mu \nu \rho \sigma}  \bar R^{\mu \nu \rho \sigma} -\bar R_{\mu \nu} \bar
R^{\mu \nu}) \end{eqnarray}
where according to eq.\ref{canon}:

\begin{eqnarray}
X=\gamma_5{\bf S}^{\mu}_{\; ;\mu}+2{\bf S}_{\mu}{\bf S}^{\mu}-
\frac{1}{4}[\gamma^{\mu},\gamma^{\nu}][d_{\mu},d_{\nu}] \nonumber \\
\bar X=\gamma_5\bar{\bf S}^{\mu}_{\; ;\mu}+
2\bar{\bf S}_{\mu}\bar{\bf S}^{\mu}-
\frac{1}{4}[\gamma^{\mu},\gamma^{\nu}][\bar d_{\mu},\bar d_{\nu}]
\end{eqnarray}

Notice that for quarks, the torsion and curvature terms are the same either
 with or without  a bar.
The explicit expression for the commutators can be
written as follows for quarks:

\begin{eqnarray}
[D_{\mu},D_{\nu}]&=&{\bf R}_{\mu \nu}+{\bf G}_{\mu \nu}+
{\bf W}_{\mu \nu}P_L+{\bf B}_{\mu \nu}+[\gamma_\mu,\gamma_\alpha]
(\frac{1}{2}\gamma_5{\bf S}^{\alpha}_{;\nu}-{\bf S}_\nu{\bf S}^{\alpha}) \\
\nonumber
&-&[\gamma_\nu,\gamma_\alpha](\frac{1}{2}\gamma_5{\bf S}^{\alpha}_{;\mu}-
{\bf S}_\mu{\bf S}^{\alpha})-{\bf S}^\alpha{\bf
S}_{\alpha}[\gamma_\nu,\gamma_\mu]
\end{eqnarray}
 and
\begin{eqnarray}
[\bar D_{\mu}, \bar D_{\nu}]&=&{\bf R}_{\mu \nu}+{\bf G}_{\mu \nu}+
{\bf W}_{\mu \nu}P_R+\bar {\bf B}_{\mu \nu}+[\gamma_\mu,\gamma_\alpha]
(\frac{1}{2}\gamma_5{\bf S}^{\alpha}_{;\nu}-{\bf S}_\nu{\bf S}^{\alpha}) \\
\nonumber
&-&[\gamma_\nu,\gamma_\alpha](\frac{1}{2}\gamma_5{\bf S}^{\alpha}_{;\mu}-
{\bf S}_\mu{\bf S}^{\alpha})-{\bf S}^\alpha{\bf
S}_{\alpha}[\gamma_\nu,\gamma_\mu]
\end{eqnarray}

For leptons we have:

\begin{eqnarray}
[D_{\mu},D_{\nu}]&=&{\bf R}_{\mu \nu}+
{\bf W}_{\mu \nu}P_L+{\bf B}_{\mu \nu}+[\gamma_\mu,\gamma_\alpha]
(\frac{1}{2}\gamma_5{\bf S}^{\alpha}_{;\nu}-{\bf S}_\nu{\bf S}^{\alpha}) \\
\nonumber
&-&[\gamma_\nu,\gamma_\alpha](\frac{1}{2}\gamma_5{\bf S}^{\alpha}_{;\mu}-
{\bf S}_\mu{\bf S}^{\alpha})-{\bf S}^\alpha{\bf
S}_{\alpha}[\gamma_\nu,\gamma_\mu]
\end{eqnarray}
and
\begin{eqnarray}
[\bar D_{\mu},\bar D_{\nu}]&=&\bar{\bf R}_{\mu \nu}+
{\bf W}_{\mu \nu}P_R+\bar {\bf B}_{\mu \nu}+[\gamma_\mu,\gamma_\alpha]
(\frac{1}{2}\gamma_5\bar{\bf S}^{\alpha}_{;\nu}-\bar{\bf S}_\nu\bar{\bf
S}^{\alpha}) \\ \nonumber
&-&[\gamma_\nu,\gamma_\alpha](\frac{1}{2}\gamma_5\bar{\bf S}^{\alpha}_{;\mu}-
\bar{\bf S}_\mu\bar{\bf S}^{\alpha})-\bar{\bf S}^\alpha\bar{\bf
S}_{\alpha}[\gamma_\nu,\gamma_\mu]
\end{eqnarray}
where we have defined for leptons:
\begin{eqnarray}
{\bf R}_{\mu\nu}=-\frac{i}{2}R^{ab}_{\;\;\;\mu\nu}\left(
\begin{array}{cc}
P_L\Sigma_{ab} & \; \\
\; & \Sigma_{ab}
\end{array}
\right)
,\; \bar{\bf R}_{\mu\nu}=-\frac{i}{2}R^{ab}_{\;\;\;\mu\nu}\left(
\begin{array}{cc}
P_R\Sigma_{ab} & \; \\
\; & \Sigma_{ab}
\end{array}
\right)
\end{eqnarray}
 and ${\bf G}_{\mu \nu}$, ${\bf W}_{\mu \nu}$ and ${\bf B}_{\mu \nu}$ are the
usual strength tensors of the corresponding gauge groups.

Once we have a consistent method for computing anomalies in a curved space-time
with torsion, let
 us
apply it to the anomalies present in the SM.

\section{Anomalies in the leptonic and baryonic currents}

In this section we will make use of the method just presented in the previous
section in order to
compute the anomalies in two global vector currents $B$ and $L$, whose
difference $B-L$ is
conserved in flat space-time
although separately
 they are not.  However, we will show that
in curved space-times the absence of right neutrinos implies that, in some
sense, gravity couples
 chirally, and thus the anomaly in the
 leptonic current acquires a gravitational contribution. Nevertheless, these
gravitational terms
 are not present in the baryonic sector, thus yielding the above commented
$B-L$ non-conservation.

Let us proceed with the computation. In order to obtain the anomalous Ward
identities related to
 the leptonic and baryonic numbers, we will consider the following local
transformations of quarks
and leptons:

\begin{eqnarray}
\psi & \rightarrow & \psi+i\alpha (x) \psi \nonumber \\
\psi^{\dagger} & \rightarrow & \psi^{\dagger} -i\psi^{\dagger} \alpha (x)
\label{tran}
\end{eqnarray}

Note that the classical action would be invariant under these transformations
if they were global .
In order to calculate
how the SM fermionic action changes, we write it in terms of a general
connection:

\begin{eqnarray}
\int d^4x \sqrt{g} {\cal L}_m=\int d^4x \sqrt{g} \frac{1}{2}(\psi^{\dagger}
 \gamma^ {\mu}D_{\mu}\psi -(D_{\mu}\psi)^{\dagger}\gamma^ {\mu} \psi)
\end{eqnarray}
where we denote by $\psi$ the leptons and quarks and $D_{\mu}$ is the gauge and
Lorentz covariant
derivative with the general connection. Under the above transformations, the
classical action
 changes as
follows:

\begin{eqnarray}
\int d^4x \sqrt{g} {\cal L}_m  \rightarrow
 \int d^4x \sqrt{g} ({\cal
L}-i\alpha(x)\nabla_{\mu}(\psi^{\dagger}\gamma^{\mu}\psi))
\label{cambio}
\end{eqnarray}
where we have integrated by parts with the Levi-Civita covariant derivative
$\nabla_{\mu}$. On the contrary, the effective action does not change under the
transformation since
it only affects to fermion fields which are integration variables:

\begin{eqnarray}
e^{-W[A,\Gamma,e]}=\int [d\psi d\psi^{\dagger}]e^{-\int d^4x \sqrt{g} {\cal
L}(\psi,\psi^{\dagger})}= \int [d\psi 'd{\psi '}^{\dagger}]e^{-\int d^4x
\sqrt{g} {\cal L}_m(\psi ',\psi '^{\dagger} )}
\label{prim}
\end{eqnarray}

Now using eqs.\ref{med} and \ref{cambio} we obtain for the effective action the
following
expression:

\begin{eqnarray}
e^{-W[A,\Gamma,e]}=\int [d\psi d\psi^{\dagger}]e^{-\int d^4x \sqrt{g}
i\alpha(x)A(x)}e^{\int d^4x \sqrt{g}
 i\alpha(x)
\nabla_{\mu}
j^{\mu}}e^{-\int d^4x \sqrt{g} {\cal L}_m(\psi,\psi^{\dagger})}
\label{sec}
\end{eqnarray}

Therefore, identifying the exponents in eqs.\ref{prim} and \ref{sec} we arrive
at:

\begin{eqnarray}
 A(x)=\nabla_{\mu}j^{\mu}
\end{eqnarray}
As we saw in the previous section the regularized expression for the anomaly is
that of eq.\ref{ar}.
 In case we applied the transformations in eq.\ref{tran} to quarks, we would
have obtained the
anomaly in the {\bf baryonic current} which is:

\begin{eqnarray}
\nabla_{\mu}j^{\mu}_B=\frac{1}{32\pi^2}\epsilon^{\mu \nu \alpha
\beta}(\frac{g^2}{2} W^a_{\mu \nu}
W^a_{\alpha\beta}+g'^2B_{\mu\nu}B_{\alpha\beta}\sum_{u,d}(y_L^2-y_R^2))
\end{eqnarray}
where the baryonic current is defined in the usual form:

\begin{eqnarray}
j^{\mu}_B=\frac{1}{N_c}Q^{\dagger}\gamma^{\mu}Q
\end{eqnarray}

We see that the result agrees with the flat space-time case. There is no
contribution
 from the
curvature nor the torsion.

 Following the same steps with the operators for leptons, we obtain the
 anomaly in
the {\bf leptonic current} which reads:

\begin{eqnarray}
\nabla_{\mu}j^{\mu}_L=\frac{1}{32\pi^2}\left \{ -\frac{\epsilon^{\alpha \beta
\gamma \delta}}{24} R_{\mu \nu \alpha \beta}
R^{\mu\nu}_{\; \; \;\gamma\delta}+\frac{\epsilon^{\alpha \beta
\gamma \delta}}{48}S_{\beta ; \gamma}S_{\delta ;
\alpha} +\epsilon^{\alpha \beta
\gamma \delta}\left(\frac{g^2}{2}W^a_{\gamma \delta}
W^a_{\alpha\beta}\right. \right. \nonumber \\
\left. \left.+g'^2B_{\gamma
\delta}B_{\alpha\beta}\sum_{\nu,e}(y_L^2-y_R^2)\right)+
\frac{1}{6} \Box S^{\alpha}_{\; ;\alpha}+ \frac{1}{96}(S^{\alpha}S^
{\nu}S_{\alpha})_{;\nu} -\frac{1}{6}(R^{\nu
\alpha}S_{\alpha}-\frac{1}{2}RS^{\nu})_{; \nu}\right\}
\end{eqnarray}
where we have defined the leptonic current as:
\begin{eqnarray}
j^{\mu}_L=L^{\dagger}\gamma^{\mu}L
\end{eqnarray}

We see in this case that, due to the non-existence of right neutrinos, some
terms depending
 on the curvature and
torsion (as total divergences) appear in the anomaly. If we had assumed their
existence , such terms would have not
 appeared and $B-L$ would be conserved, as it happens in flat space time,
provided the following
relation is satisfied:

\begin{eqnarray}
\sum_{u,d}(y_L^2-y_R^2)=\sum_{\nu,e}(y_L^2-y_R^2)
\end{eqnarray}
which is indeed the case for the usual SM hypercharge assignment.

\section{Gauge anomalies}

In the previous section we have shown that the formulation of the SM in a
curved space-time
with torsion may drive to the non-conservation of  global currents $B-L$ which
however are
 conserved in
flat space-time. In this section we will study whether something similar
happens to gauge currents.
The non-conservation of gauge currents would destroy the consistency of the
model. On the
other hand one may wonder, whether the cancellation of new contributions due to
curvature and
 torsion
could impose new constraints to hypercharge assignments.

Let us begin by writing the effective action for the SM matter sector :

\begin{eqnarray}
e^{-W[A,\Gamma,e]}=\int [d\psi d\psi^{\dagger}]e^{-\int d^4x\sqrt{g}{\cal L}_m}
\label{ea}
\end{eqnarray}

The matter lagrangian given in eq.\ref{lm} is invariant under the
$SU(3)_c\times
 SU(2)_L\times U(1)_Y$ gauge transformations, which are given by:

\begin{eqnarray}
Q & \rightarrow & Q-i\theta^a(x)\Lambda^a Q\nonumber \\
Q^{\dagger} & \rightarrow & Q^{\dagger}+i\theta^a(x)Q^{\dagger}\Lambda^a
\end{eqnarray}
where, following the definitions at the beginning of the this chapter,
$\Lambda^a$ are the
 $SU(3)_c$ generators in the appropriate representation. These
transformations only affect to quarks since they are the only fields that
couple to $SU(3)_c$. We
 also have:

\begin{eqnarray}
\psi & \rightarrow & \psi-i\theta^a(x)T^a P_L\psi \nonumber \\
\psi^{\dagger} & \rightarrow & \psi^{\dagger} +i\theta^a(x)\psi^{\dagger}P_R
T^a
\end{eqnarray}
Here $T^a$ are the $SU(2)_L$ generators in the appropriate representation and
$\psi$ stands for
 quarks
or leptons. Finally, the hypercharge transformation reads

\begin{eqnarray}
\psi & \rightarrow & \psi-i\theta (x)({\bf y_L} P_L+{\bf y_R}P_R)\psi \nonumber
\\
 \psi^{\dagger} & \rightarrow & \psi^{\dagger}+i\theta (x)\psi^{\dagger}({\bf
y_L}P_R+{\bf y_R}P_L)  \end{eqnarray}
where ${\bf y_L}$ and ${\bf y_R}$ are the diagonal hypercharge matrices in
flavor
space which appear
 in eq.\ref{hy}.
As it is well known, in spite of the invariance of the lagrangian under the
above gauge
transformations, the effective action may have an anomalous variation due to
the integration measure.
In the  following we will obtain the expression for the anomalous variation of
the effective action
in the
 case of
$SU(3)_C$ transformations, but the result will be equally valid for the other
groups. Let us
first introduce the notation  ${\bf\theta}=-i\theta^a\Lambda^a$,
$D_{\mu}{\bf\theta}= \partial_{\mu}\theta+
[{\bf G}_{\mu},{\bf\theta}]$. We will use that:

\begin{eqnarray}
\frac{\delta W}{\delta G_{\mu}^a}=-ig_s\langle Q^{\dagger}\gamma^{\mu}\Lambda^a
Q \rangle=
-ig_s\langle j^{\mu a}\rangle
\end{eqnarray}
is the expectation value of the gauge current in presence of the background
fields. We will also define:
\begin{eqnarray}
\langle j^{\mu} \rangle=\langle j^{\mu a} \Lambda ^a \rangle
\end{eqnarray}

Under the previously mentioned $SU(3)_C$ transformations the gauge fields
change as follows:

\begin{eqnarray}
{\bf G}_{\mu}\rightarrow {\bf G}_{\mu}-D_{\mu}{\bf \theta}
\end{eqnarray}
or in components:
\begin{eqnarray}
G_{\mu}^c\rightarrow G_{\mu}^c-\frac{1}{g_s}
\partial_{\mu}\theta^c+G_{\mu}^b\theta^a f^{abc}
\end{eqnarray}
and the anomalous change in the effective action is given by:

\begin{eqnarray}
 W[G-D\theta,\Gamma,e]- W[G,\Gamma,e] & = &  -\int d^4x \sqrt{g}\left (
\frac{1}{g_s}
 \partial_{\mu}\theta^c+G_{\mu}^a\theta^b f^{abc}\right )\frac{\delta W}{\delta
G_{\mu}^c} \nonumber \\
   &=&  \int d^4x \sqrt{g}\theta^b\left ( \frac{1}{g_s}\nabla_{\mu}\frac{\delta
W}{\delta G_{\mu}^b}+
G_{\mu}^a\frac{\delta W}{\delta G_{\mu}^c}f^{acb}\right ) \nonumber \\
 & = & -\int d^4x \sqrt{g}i\theta^b(D_{\mu}\langle j^{\mu} \rangle)^b
\label{caea}
\end{eqnarray}
where we have integrated by parts with the Levi-Civita covariant derivative and
we have made use of the symmetry properties of the structure constants
$f^{abc}$.
Notice that we have denoted:
 \begin{eqnarray}
D_{\mu}\langle j^{\mu} \rangle=\nabla_{\mu}\langle j^{\mu} \rangle
+[{\bf G}_{\mu},\langle j^{\mu} \rangle]
\end{eqnarray}
and:
\begin{eqnarray}
 D_{\mu}\langle j^{\mu} \rangle=(D_{\mu}\langle j^{\mu} \rangle)^a\Lambda^a
\end{eqnarray}

The change in the integration measure can be computed in the standard fashion
as we
did in the abelian case and it yields:
\begin{eqnarray}
[dQ' dQ'^{\dagger}]= [dQ dQ^{\dagger}]e^{i\sum_{n}\int d^4x\sqrt{g}
(\phi^{\dagger}_n\theta^a(x)\Lambda^a\phi_n-
\xi^{\dagger}_n\theta^a(x)\Lambda^a\xi_n)}
\end{eqnarray}
where $\phi_n$ and $\xi_n$ are given in eq.\ref{aut}. The anomaly is defined
as:
\begin{eqnarray}
A^a(x)_{SU(3)}=\sum_{n}(\phi_n^{\dagger}\Lambda^a \phi_n
-\xi_n^{\dagger} \Lambda^a\xi_n)
\end{eqnarray}

Therefore we can write the transformed effective action as

\begin{eqnarray}
e^{-W[G',\Gamma,e]}=\int [d\psi d\psi^{\dagger}]e^{-\int d^4x \sqrt{g} {\cal
L}_m}e^{-i\int d^4x \sqrt{g}
\theta^a(x)A^a(x)}
\end{eqnarray}
 Expanding to first order in $\theta$ and identifying with eq.\ref{caea} we
obtain:

\begin{eqnarray}
(D_{\mu}\langle j^{\mu} \rangle)^a=A^a(x)
\end{eqnarray}
This anomalous Ward identity implies that the non-conservation of the gauge
current expectation value is given by the anomaly coefficient. Finally the
expression
 for the anomaly in  the $SU(2)_L$ and
$U(1)_Y$ currents can be computed in the same way and are  given by:

\begin{eqnarray}
A^a(x)_{SU(2)} & = &
\sum_{n}(\phi_n^{\dagger}P_LT^a\phi_n-\xi_n^{\dagger}P_RT^a\xi_n) \nonumber \\
A(x)_{U(1)} & = & \sum_{n}(\phi_n^{\dagger}({\bf y_L}P_L+{\bf
y_R}P_R)\phi_n-\xi_n^{\dagger}({\bf
y_L}P_R+{\bf y_R}P_L) \xi_n)
\end{eqnarray}

As we have said before these expressions for the anomalies need regularization
and, as we did in
 the abelian case,
we will use the operators $H_{\psi}$ and $H_{\psi^{\dagger}}$ defined in
eq.\ref{op}
to regularize each piece of the anomaly separately. The results are the
following:\\

{\bf Anomaly in the $SU(3)_c$ gauge current.}
\begin{eqnarray}
A^a_{SU(3)}(x)=\frac{1}{(4\pi)^2}tr(\Lambda^a(a_2(H_{\psi},x)-a_2(H_{\psi^{\dagger}},x)))
 \end{eqnarray}
which for the divergence of the current gives:

\begin{eqnarray}
(D_{\mu}\langle j^{\mu }\rangle)^a=-\frac{1}{32\pi^2}g_sg'\epsilon^{\mu \nu
\alpha \beta}G^a_{\mu
\nu} B_{\alpha\beta}\sum_{u,d}(y_L-y_R)
\end{eqnarray}

This result agrees with that found in flat space-time. There are no new
contributions from
 curvature or
torsion. The only term present depends on the strength tensor of the
hypercharge fields.
 The cancellation condition for this anomaly is given by:

\begin{eqnarray}
\sum_{u,d}(y_L-y_R)=0
\label{c1}
\end{eqnarray}
\\

{\bf Anomaly in the $SU(2)_L$ gauge current}

Following the same steps as before for the $SU(2)_L$ transformations, we find:
\begin{eqnarray}
A^a_{SU(2)}(x)=\frac{1}{(4\pi)^2}tr(T^a(a_2(H_{\psi},x)P_L-a_2(H_{\psi^{\dagger}},x)P_R))
 \end{eqnarray}

The expression for the divergence of the gauge current can be obtained after
some algebra and
 it yields:
\begin{eqnarray}
 (D_{\mu}\langle j^{\mu}\rangle)^a=-\frac{1}{32\pi^2}gg'\epsilon^{\mu \nu
\alpha \beta}W^a_{\mu \nu}
B_{\alpha\beta}(\sum_{u,d}N_Cy_L+\sum_{\nu,e}y_L)
\end{eqnarray}

We observe that the result is the same as in flat space-time. All the
contributions coming from
 curvature
or torsion vanish, and the only term present depends on the strength fields of
$SU(2)_L$ and the
 hypercharge
field. The cancellation condition reads in this case:
\begin{eqnarray}
 \sum_{u,d}N_Cy_L+\sum_{\nu,e}y_L=0
\label{c2}
\end{eqnarray}
\\

{\bf Anomaly in the $U(1)_Y$ gauge current}

Finally, the expression for the anomaly in the $U(1)_Y$ current can be written
as:
\begin{eqnarray}
A_{U(1)}(x)=\frac{1}{(4\pi)^2}tr(({\bf y_L}P_L+{\bf
y_R}P_R)a_2(H_{\psi},x)-({\bf y_L}P_R+ {\bf y_R}P_L)
a_2(H_{\psi^{\dagger}},x))
\end{eqnarray}
where  ${\bf y_L}$ and ${\bf y_R}$ are the hypercharge matrices. The final
expression for the
 divergence
of the gauge current is now more involved than the non-abelian cases due to the
appearance
of terms depending on curvature and torsion. The result goes as follows:

\begin{eqnarray}
D_{\mu}\langle j^{\mu} \rangle & =
&\frac{1}{32\pi^2}\left(\left[
\sum_{u,d}N_C(y_L-y_R)+\sum_{\nu,e}(y_L-y_R)\right ]
\left ( -\frac{1}{24}\epsilon^{\alpha \beta \gamma \delta} R_{\mu \nu \alpha
\beta} R^{\mu \nu}_{\; \; \gamma\delta} +\frac{1}{6}\Box S^{\mu}_{\; ;
\mu} \right. \right.\nonumber \\
  & + &\left. \frac{1}{96}(S^{\alpha}S^{\nu}S_{\alpha})_{;\nu}+\frac{1}{48}
\epsilon^{\alpha \beta \gamma \delta}S_{\beta ; \gamma}S_{\delta ; \alpha}-
\frac{1}{6}
(R^{\nu \alpha}S_{\alpha}-\frac{1}{2}RS^{\nu})_{; \nu}\right ) \nonumber \\
 & + & \frac{g_s^2}{2}\epsilon^{\mu \nu \alpha \beta}G^a_{\mu \nu}G^a_{\alpha
\beta}\sum_{u,d}(y_L-y_R)+ \frac{g^2}{4}\epsilon^{\mu \nu \alpha \beta}W^a_{\mu
\nu}W^a_{\alpha \beta}(\sum_{u,d}N_Cy_L+ \sum_{\nu,e}y_L) \nonumber \\
 & + & \left. {g'}^ 2 \epsilon^{\mu \nu \alpha \beta}B_{\mu
\nu}B_{\alpha \beta}
(\sum_{u,d}N_C(y_L^3-y_R^3)+\sum_{\nu,e}(y_L^3-y_R^3))\right )
\end{eqnarray}

Notice the appearance of terms depending on curvature and torsion which did not
occur in
 the case of
non-abelian gauge fields although they are also chiral.
The new terms that were not present in flat space-time  impose a new
cancellation condition: the
 vanishing of the sum of all hypercharges:
\begin{eqnarray}
\sum_{u,d}N_C(y_L-y_R)+\sum_{\nu,e}(y_L-y_R)=0
\label{c3}
\end{eqnarray}
on the other hand we have  that the cancellation of the terms already present
in flat space
 time gives the conditions:

\begin{eqnarray}
0 & = & \sum_{u,d}(y_L-y_R)  \label{c4} \\
0 & = & \sum_{u,d}N_Cy_L+\sum_{\nu,e}y_L
\label{c5} \\
0 & = & \sum_{u,d}N_C(y_L^3-y_R^3)+\sum_{\nu,e}(y_L^3-y_R^3)
\label{c6}
\end{eqnarray}

The conditions in eqs.\ref{c4} and \ref{c5} are respectively the same as those
in eqs.\ref{c1}
 and \ref{c2}, therefore there are four independent conditions and five
hypercharges, namely:
$y_L^{\nu}=y_L^{e}, \;y_L^{u}=y_L^{d},\; y_R^{e},\; y_R^{u}\; ,y_R^{d}$.

\section{Gravitational anomalies}

As we have mentioned above, the EP  states that any theory in curved space-time
should be invariant under local Lorentz transformations. In this section, we
consider the possible
violation of this local symmetry due to quantum effects when chiral fermions
are present
\cite{chang}, as indeed
 happens in
the SM. We will conclude that whenever abelian chiral gauge fields are present,
as it is
the case of the hypercharge field, local Lorentz invariance is violated.
However, due to the specific
hypercharge assignment in the SM this anomaly is exactly cancelled. The
condition for the
cancellation of the Lorentz anomaly is the same as that of the cancellation of
terms depending on
curvature and torsion in the $U(1)_Y$ anomaly eq.\ref{c3}.

Let us proceed with the explicit computation. Under local Lorentz
transformations the spinor,
 vierbein and
connection fields present in the matter lagrangian of the SM, eq.\ref{lm},
change as:

\begin{eqnarray}
\psi(p) & \rightarrow & e^{\frac{i}{2}\epsilon^{mn}(x)\Sigma_{mn}}\psi(p)
\nonumber \\
\psi^{\dagger}(p) & \rightarrow &
\psi^{\dagger}(p)e^{-\frac{i}{2}\epsilon^{mn}(x)\Sigma_{mn}}
\nonumber \\ e^a_{\;\mu} & \rightarrow & e^a_{\;\mu}-\epsilon ^a_{\;
b}(x)e^b_{\;\mu} \nonumber \\
\Gamma^{a \; b}_{\; \mu} & \rightarrow & \Gamma^{a \; b}_{\; \mu}+\epsilon
^a_{\; c}(x)\Gamma^
{c \; b}_{\; \mu}-
\epsilon _c^{\; b}(x)\Gamma^{a \; c}_{\; \mu}-\partial_{\mu}\epsilon^{ab}(x)
\end{eqnarray}

Under this set of transformations the matter lagrangian is invariant. However,
it may happen that
 the
effective action in eq.\ref{ea} suffers an anomalous change as it occurs for
gauge fields. This
 change is
given by:

\begin{equation}
\begin{array}{l}
W[A,\Gamma-D\epsilon, e-\epsilon e]=  \\
 W[A,\Gamma,e]
 -  \displaystyle\int d^4x \sqrt{g} \left( (-\epsilon ^a_{\; c}(x)\Gamma^{c \;
b}_{\; \mu}+ \epsilon _c^{\; b}(x)\Gamma^{a \; c}_{\;
\mu}+\partial_{\mu}\epsilon^{ab}(x))\frac{\delta W} {\delta \Gamma ^{a\; b}_{\;
\mu}}
  +  \epsilon^a_{\; b}(x)e^b_{\; \mu}\frac{\delta W}{\delta e^a_{\; \mu}}
\right )
\end{array}
\end{equation}
Here we have denoted by $A$ all the gauge fields in the theory. Now, if we
integrate by parts
 and use the antisymmetry of the connection components $\Gamma^{a\; b}_{\;\mu}$
in $a$ and $b$,  we
can rewrite this expression as

\begin{equation}
\begin{array}{l}
W[A,\Gamma-D\epsilon, e-\epsilon e]=  \\
W[A,\Gamma,e]+\displaystyle\int d^4x\sqrt{g}\epsilon^{ab}\left( \nabla_{\mu}
\frac{\delta W}{\delta \Gamma^{a\; b}_{\; {\mu}}}+\Gamma_{a\mu}^{\; \;
c}\frac{\delta W}{\delta
\Gamma^{c \; b}_{\; \mu}}-\Gamma^c_{\; \mu b}\frac{\delta W}{\delta
\Gamma^{a \; c}_{\; \mu}}-T_{ab}\right)
\end{array}
\end{equation}
where $T_{ab}=e^b_{\; \mu}{\delta W}/\delta e^a_{\; \mu}$ is the
expectation value of the energy-momentum tensor in presence of the background
 fields. We can write this result more conveniently
using the following definitions:

\begin{eqnarray}
\frac{\delta W}{\delta \Gamma^{a\; b}_{\; \mu}} & = & -\frac{i}{4}\langle
\psi^{\dagger}(\gamma^{\mu}\Sigma^{ab}+ \Sigma^{ab} \gamma^{\mu})\psi \rangle
=
-\frac{i}{2}\langle j^{ab}_{\; \; \nu} \rangle \nonumber \\
\langle j_{\mu} \rangle & = & \langle
j^{ab}_{\; \; \nu} \Sigma_{ab} \rangle \nonumber \\
D_{\mu}\langle j^{\mu} \rangle  & = & \nabla_{\mu}
\langle j^{\mu} \rangle +[\Gamma_{\mu},\langle j^{\mu} \rangle ] \\ \nonumber
\Gamma_{\mu} &  = & \Gamma^{a \; b}_{\; \mu} \Sigma_{ab}
\end{eqnarray}

Therefore we can rewrite:

\begin{eqnarray}
 W[A,\Gamma-D\epsilon, e-\epsilon e]=W[A,\Gamma,e]+\int
d^4x\sqrt{g}\epsilon_{ab}(x)(-\frac{i}{2}
(D_{\mu}\langle j_{\mu} \rangle)^{ab}-T^{ab})
\label{eac}
\end{eqnarray}

In addition, we can calculate the change in the effective action due to the
change in the
integration measure as we did for the gauge anomaly and obtain:
\begin{eqnarray}
e^{-W[A,\Gamma',e']}=\int [d\psi d\psi^{\dagger}]e^{-\int d^4x \sqrt{g}
{\cal L}_m}e^{-\frac{i}{2}\int d^4x
\sqrt{g}(\epsilon_{ab}(x)A^{ab}(x))}
\label{ca}
\end{eqnarray}
where:
\begin{eqnarray}
A^{ab}(x)=\sum_{n}\int d^4x \sqrt{g}
(\phi_n^{\dagger}\Sigma^{ab}\phi_n-\xi_n^{\dagger}\Sigma^{ab}\xi_n)
\label{al}
\end{eqnarray}

Finally, expanding eq.\ref{ca} to first order in $\epsilon$ and  identifying
the terms
 in eq.\ref{eac}, we find
the anomalous identity:

\begin{eqnarray}
A^{ab}(x)=-(D_{\mu}\langle j^{\mu} \rangle)^{ab}+i(T^{ab}-T^{ba})
\end{eqnarray}

The expression for the anomaly in eq.\ref{al} needs regularization. As we did
in all the previous
 cases
we use the operators $H_{\psi}$ and $H_{\psi^{\dagger}}$ to regularize the
first and second terms
 respectively.
The result can be expressed as follows:

\begin{eqnarray}
A^{mn}_{SO(4)}(x)=\frac{1}{(4\pi)^2}tr(\Sigma^{mn}(a_2(H_{\psi},x)-
a_2(H_{\psi^{\dagger}},x)))
 \end{eqnarray}

After a lengthy calculation we arrive to the final expression for the Lorentz
anomaly:

\begin{eqnarray}
A^{mn}(x) & = & \frac{g'}{32\pi^2}\left (\frac{1}{6}\epsilon^{mnab}R_{\mu \nu a
b}B^{\mu \nu}-\frac{1}{6} (B^n_{\; \;\nu}S^{m;\nu}-B^m_{\;\; \nu}S^{n;\nu})
\right. \nonumber \\
 & - &
\frac{1}{24}\epsilon^{mnab}(B_{ac}S^cS_b+B_{ab}S^2)-\frac{1}{6}\epsilon^{mnab}B_{ab}R
 -\frac{1}{2}S^{\mu}_{\; ;\mu}B^{mn}\nonumber \\
 & - & \left. \frac{1}{3}\epsilon^{mnab}\Box B_{ab})(\sum_{u,d}N_c(y_L-y_R)+
\sum_{\nu,e}(y_L-y_R)\right)
\label{lal}
\end{eqnarray}

Notice that pure gravity terms do not occur. Indeed it has been shown that
there are no pure
 gravitational
anomalies in four dimensions \cite{alwi}. Observe also that all the terms
depend on the $B_{ab}$
field, which
 is abelian,
whereas there is no contribution from non abelian gauge fields. Finally, the
cancellation
 condition agrees with that of
eq.\ref{c3} which ensures the vanishing of the gravity terms in the $U(1)_Y$
anomaly and, as we
 have  already commented, is satisfied in the SM. It is also interesting to
realize that eq.\ref{lal}
without curvature and  torsion terms reduces to:
\begin{eqnarray}
A^{mn}(x)=\frac{g'}{32\pi^2}(-\frac{1}{3}\epsilon^{mnab}\Box B_{ab})
\left (\sum_{u,d}N_c(y_L-y_R)+ \sum_{\nu,e}(y_L-y_R)\right)
\label{plano}
\end{eqnarray}

This is the expression for the anomaly in the global Lorentz current in flat
space-time, which
classically is a basic symmetry in any relativistic quantum field theory.
 Notice again that the specific hypercharge assignment in the SM  allows its
cancellation. The last remark is important since, the cancellation condition in
eq.\ref{c3} or
eq.\ref{lal} were obtained in curved space-time and are also referred as mixed
gauge-gravitational
anomalies. However eq.\ref{plano} is obtained in flat space-time. Therefore, in
flat space-time
without any reference to gravitation, it is also possible to obtain the four
anomaly cancellation
conditions mentioned before.

\section{Charge quantization in the Standard Model}

In this section we will discuss the consequences of the requirement of the
cancellation of the above computed gauge and gravitational anomalies.
The set of eqs.\ref{c1}, \ref{c2}, \ref{c3} for the vanishing of gauge
anomalies and
eq.\ref{plano} for the absence of global Lorentz anomalies in each family
provide four equations  for the five unknowns $y_L^{\nu}=y_L^{e},
\;y_L^{u}=y_L^{d},\; y_R^{e},\; y_R^{u}\; ,y_R^{d}$. Let us try to solve the
system explicitly and accordingly to check whether they fix all the
hypercharges
up to a normalization factor \cite{iba}. First, we note that the four equations
can be reduced to just one
 equation
 for
two unknowns, namely:

\begin{eqnarray}
21y^{u2}_Ry^d_R+21y^{u2}_Ry^d_R+6y^{u3}_R+6y^{d3}_R=0
\end{eqnarray}
This equation, in turn, can be expressed in terms of one variable
for $y^d_R \ne 0$:

\begin{eqnarray}
1+\left(\frac{y^u_R}{y^d_R}\right)^3+\frac{21}{6}\left(
\frac{y^u_R}{y^d_R}\right)^2+\frac{21}{6}\frac{y^u_R}{y^d_R}=0
 \end{eqnarray}

Now it is not difficult to see that there are three real solutions for this
equation:  \begin{eqnarray}
\frac{y^u_R}{y^d_R}=-1,\; -2,\; -\frac{1}{2}
\end{eqnarray}

The quotient determines the rest of hypercharges as follows:

\begin{eqnarray}
y^u_L & = & y^d_L=\frac{1}{2}(y^u_R+y^d_R),\\ \nonumber
y^e_L & = & y^{\nu}_L=-\frac{3}{2}(y^u_R+y^d_R) \\ \nonumber
y^e_R  & = & -3(y^u_R+y^d_R)
\end{eqnarray}

Therefore, there are three possible sets of hypercharges (up to a normalization
factor) which
explicitly
 read:

\begin{eqnarray}
y^u_R & = & -y^d_R \\ \nonumber
y^u_L & = & y^d_L=y^e_L=y^{\nu}_L=y^e_R=0,
\end{eqnarray}

\begin{eqnarray}
y^u_R & = & -2y^d_R\\ \nonumber
y^u_L & = & y^d_L=-\frac{1}{2}y^d_R \\ \nonumber
 y^e_L & = & y^{\nu}_L= \frac{3}{2}y^d_R\\ \nonumber
 y^e_R & = & 3y^d_R,
\end{eqnarray}
and
\begin{eqnarray}
y^d_R & = & -2y^u_R\\ \nonumber
y^u_L & = & y^d_L=-\frac{1}{2}y^u_R \\ \nonumber
 y^e_L & = & y^{\nu}_L= \frac{3}{2}y^u_R\\ \nonumber
 y^e_R & = & 3y^u_R
\end{eqnarray}

The second set provides the usual assignment in the SM. The first one is the
"bizarre"  hypercharge assignment obtained in \cite{Ramond2},
and finally the third one can be obtained from the usual one by exchanging the
hypercharges
of $u$ and $d$ quarks. With the standard weak isospin assignment, the last set
leads to  different electric charges for the left and right components of the
quark fields and therefore to
 chiral electromagnetism. To summarize, anomaly cancellation arguments alone do
not
 fix the hypercharges in the SM; further physical constraints as the vector
character of the
electromagnetism and the existence of charged electrons are needed for that
purpose.

\section{Conclusions}

In this work we have carefully computed the different anomalies that appear in
the Standard
Model (SM) defined in a classical background space-time with torsion.

The Equivalence Principle can be used to completely define the nature of the
coupling of the
Standard Model fields to the vierbein and the metric connection. In particular
only fermions need to be minimally coupled to the torsion. The addition of
other
non minimal couplings give rise in some cases to gauge anomalies that cannot be
cancelled by any hypercharge assignment.

Concerning the anomalies affecting global currents we arrive to the following
results. The
baryonic current anomaly is not modified by any curvature or torsion term and
then it is the
same as in flat space-time. However, due to the absence of the right-handed
neutrinos, the leptonic current anomaly gets new contributions coming from
curvature and torsion terms. Therefore, the conservation of the total baryonic
minus leptonic number ($B-L$), which is known to apply for the SM in flat
space-time, is violated when curvature and torsion are present. This fact could
be relevant in connection with the problem of the baryonic asymmetry of the
Universe.

The gauge anomalies corresponding to the groups $SU(3)_C$ and $SU(2)_L$ do not
get new
contributions and then we find the standard conditions for their cancellation
in
terms of the fermion hypercharges. For the $U(1)_Y$ anomaly we obtain
contributions from all the SM gauge fields and also new curvature and torsion
terms. The cancellation of these gauge and gravitational terms gives rise to
two
new conditions on the hypercharge assignments in addition to the other two
mentioned above.

The gravitational anomaly has been computed as a gauge anomaly corresponding to
the local
(Euclidean)  Lorentz group $SO(4)$. This anomaly has contributions from terms
which are products of the hypercharge gauge field and curvature (mixed
gauge-gravitational anomalies), hypercharge and torsion  and hypercharge alone.
This is consistent with the well known fact of the absence of  pure
gravitational anomalies in four dimensions. On the other hand, the only
condition on the hypercharges found to cancel this terms is exactly the same
found to cancel the curvature and the torsion terms appearing in the $U(1)_Y$
gauge anomaly. At this point we would like to stress that, even when the
curvature and the torsion vanish, we find a term contributing to this anomaly
which depends only on the $U(1)_Y$ field. Therefore, the corresponding anomaly
equation is just the same that one find when computing the anomaly of the
global
current corresponding to the Lorentz group for the SM in flat space-time.
Therefore, the condition on the hypercharge assignment that is usually referred
as
coming from the cancellation of the mixed gauge-gravitational anomaly in the
SM,
can be obtained without any reference to  gravitation just by demanding the
cancellation of the global Lorentz anomaly in the SM. In spite of the fact that
this condition does not correspond to the cancellation of a gauge  but a global
anomaly (which in principle does not destroy the consistency of the theory), it
is quite natural to be required since it amounts to the preservation of the
Special Relativity Principle at the quantum level.

Finally we have deal with the problem of the hypercharge assignments (family by
family) from the cancellation of the gauge and mixed gauge-gravitational
anomalies (or global Lorentz anomalies according to the discussion above) in
the
SM. In principle we have four equations and five unknowns. If the solution to
this equations were unique one could fix the hypercharges modulo the global
normalization that could be determined for example from the electron charge
(provided it were different from zero). However, as a result of our analysis we
find three independent kind of solutions. One is the .bizarre. solution found
in
\cite{Ramond2}. The second is the one that contains the standard hypercharge
assignment of the SM and then leads to the (fractional) quantization of the
electric charge. The third  one is analogous to the second one but exchanging
the $u$ quark hypercharge by the $d$ quark one. From the phenomenological point
of view only the second solution is acceptable since the first one produce a
chargeless electron and the third one chiral electromagnetism. Those are the
facts and we consider a matter of personal taste to decide if the cancellation
of
the anomalies determines or almost determines the SM hypercharges and the
(fractional) quantization of the electric charges in the SM.   In any case we
would like to remark again that, as discussed above, the four hypercharge
conditions can be obtained entirely in the context of the SM in flat
space without the introduction of gravitation at all.

 {\bf Acknowledgements:}
This work has been partially supported by the Ministerio de Educaci\'on y
Ciencia (Spain)(CICYT AEN93-0776).

\end{document}